\documentclass[12pt,epsfig]{article}
\makeatletter
\def\vereq#1#2{\lower3pt\vbox{\baselineskip1.5pt \lineskip1.5pt
\ialign{$\m@th#1\hfill##\hfil$\crcr#2\crcr\sim\crcr}}}
\makeatother
\newcounter{axn}

\def\diag{{\rm diag}} 
\newcommand{\bequ}{\begin{equation}}
\newcommand{\eequ}{\end{equation}}
\newcommand{\beqn}{\begin{eqnarray}}
\newcommand{\eeqn}{\end{eqnarray}}
\newcommand{\bctr}{\begin{center}}
\newcommand{\ectr}{\end{center}}
\newcommand{\Ls}{\left(}
\newcommand{\Rs}{\right)}

\newcommand{\Ll}{\left[}
\newcommand{\Rl}{\right]}
\newcommand{\LL}{\left.}
\newcommand{\RR}{\right.}
\newcommand{\hsp}[1]{\hspace {#1cm}}

\newcommand{\nn}{\nonumber}
\newcommand{\VEV}[1]{\left\langle #1 \right\rangle}
 
\newcommand{\eff}{{\rm {eff}}}
\def\PR#1#2#3{Phys. Rev. {\bf #1} (#3) #2 }
\def\PRL#1#2#3{Phys. Rev. Lett. {\bf #1} (#3) #2 }
\def\PL#1#2#3{Phys. Lett. {\bf #1} (#3) #2 }

\def\NP#1#2#3{Nucl. Phys. {\bf #1} (#3) #2 }

\def\PTP#1#2#3{Prog. Theor. Phys. {\bf #1} (#3)#2 }

\begin{document}

\begin{titlepage}
\begin{flushright}
hep-ph/0403264\\
KUNS-1904 
\end{flushright}

\vskip 1cm
\begin{center}
{\large\bf Vacuum structure in 5D $SO(10)$ GUT on $S^1/Z_2$}

\vspace{0.5cm}

{\bf  Naoyuki Haba$^{1}$ and Toshifumi Yamashita$^{2}$}\\ 
\vspace{0.5cm}
$^1${\small \it Institute of Theoretical Physics, 
 University of Tokushima, Tokushima 770-8502, Japan} \\
$^2${\small \it Department of Physics, Kyoto University,
Kyoto, 606-8502, Japan}\\
\end{center}
\vskip .5cm
\vspace{.3cm}

\begin{abstract}

We study the vacuum structure in 5D 
 $SO(10)$ grand unified theory (GUT) 
 compactified on $S^1/Z_2$ orbifold, 
 where $SO(10)$ is broken into 
 $SU(3)\times SU(2)\times U(1)^2$ 
 through the boundary conditions. 
Although a lot of people extended to 6D 
 to avoid massless colored particle, 
 we can show they obtain
 finite masses 
 by the radiative corrections. 
In a supersymmetric case, 
 the fermionic partner of the zero-mode 
 can also acquire non-vanishing mass 
 through the SUSY breaking effects,  
 and the gauge coupling unification
 can be recovered by use of brane localized 
 kinetic terms.

\end{abstract}
\end{titlepage}
\setcounter{footnote}{0}
\setcounter{page}{1}
\setcounter{section}{0}
\setcounter{subsection}{0}
\setcounter{subsubsection}{0}

\section{Introduction}

Grand unified theories (GUTs) are very attractive models 
 in which the three
 gauge groups are unified at a high energy scale. 
However, one of the most serious problems 
 to construct a model of GUTs 
 is how to realize the mass splitting  between the triplet and 
 the doublet Higgs particles  in the Higgs sector. 
This problem is so-called  triplet-doublet (TD) splitting problem. 
A new idea for solving the TD splitting problem 
 has been suggested in higher dimensional 
 GUTs where the extra dimensional 
 coordinates are compactified on 
 orbifolds\cite{5d}-\cite{so(10)others}. 
In these scenarios, 
 Higgs and gauge fields are propagating in
 extra dimensions, and  
 the orbifolding realizes the gauge group
 reduction and 
 the TD splitting since  
 the doublet (triplet) Higgs fields have (not)
 Kaluza-Klein (KK)\cite{KK} zero-modes. 
A lot of attempts and progresses have been done in 
 the extra dimensional GUTs on orbifolds%
 \cite{6d}-\cite{quasilocalize}.
Among them, the reduction of $SO(10)$ gauge symmetry 
 and the TD splitting solution are 
 first considered in 6D models in Refs.\cite{ABC} and \cite{HNOS}, 
 where $SO(10)$ is broken into 
 $SU(3)\times SU(2)\times U(1)^2$ 
 through the boundary conditions. 
5D $SO(10)$ models have been also considered, for example, 
 in Refs.\cite{5dso10shafi} and \cite{5dso10ps}. 
However, 
 in Ref.\cite{5dso10shafi}, 
 $SO(10)$ is reduced into 
 $SU(3)\times SU(2)\times U(1)^2$,  
 but some colored fields are 
 remaining as the zero-modes of 
 the extra dimensional components of 
 the gauge fields. 
The introduction of 
 a pair of bulk $\bf{16}$ multiplets and a bulk singlet 
 that develop non-vanishing vacuum expectation values (VEVs) 
 can make the zero-modes massive, 
 the model became a little complicated. 
On the other hand, 
 in Ref.\cite{5dso10ps}, 
 $SO(10)$ is broken to 
 only the Pati-Salam $SU(4)\times SU(2) \times SU(2)$ 
 symmetry\cite{PS} by the boundary condition, 
 where 
 all the extra dimensional components of 
 the gauge fields acquire masses of the order 
 compactification scale.
This Pati-Salam gauge symmetry should be 
 broken down to the standard gauge
 symmetry by the usual Higgs mechanism.

In this paper we reanalyze the 5D theory 
 on a $S^1/Z_2$ orbifold where
 $SO(10)$ is broken into 
 $SU(3)\times SU(2)\times U(1)^2$ 
 by the boundary conditions. 
Colored fields of the Wilson line degrees of
 freedom are surely remaining
 as the zero-modes in the tree level. 
However, we will show they obtain
 finite masses of order the compactification scale 
 by the radiative corrections.
In the supersymmetric (SUSY) case, 
 the fermionic partners of the zero-modes 
 can also acquire non-vanishing masses 
 through the SUSY breaking effects. 
Although the scale of the masses is the SUSY breaking scale 
 and the gauge coupling unification seems to be disturbed,
 the suitable boundary localized kinetic terms may 
 conserve the gauge coupling unification.

\section{$SO(10)$ GUT on orbifold}

We consider a 5D $SO(10)$ GUT in which  
 the gauge and Higgs fields 
 propagate in the bulk. 
The 5th dimensional coordinate $(y)$ 
 is assumed to be compactified on an $S^1/Z_2$ 
 orbifold. 
Under the parity transformation of $Z_2$ 
 which transforms $y \rightarrow -y$,  
 the gauge field $A_M(x^\mu,y)$ $(M= \mu( =0$-$3),5)$ 
 in 5D 
 transforms as 
\begin{eqnarray}
 && A_\mu(x^\mu,-y) = PA_\mu(x^\mu,y)P^{\dagger},\\
 && A_5(x^\mu,-y) = -PA_5(x^\mu,y)P^{\dagger},
\end{eqnarray}
where $P$ is the operator of 
 $Z_2$ transformation. 
Two walls at $y=0$ and $\pi R$ are
 fixed points under $Z_2$ transformation. 
The physical space can be taken to $0 \leq y \leq \pi R$. 
Considering the $S^1$ 
 boundary condition, 
 $A_M(x^\mu,y+2\pi R) = TA_M(x^\mu,y)T^{\dagger}$, 
 the reflection 
 around $y=\pi R$, $Z_2'$,   
 is given by $P'=TP$. 
The gauge field $A_M(x^\mu,y)$ 
 transforms 
\begin{eqnarray}
 && A_\mu(x^\mu,\pi R-y) = P'A_\mu(x^\mu, \pi R+y)P'^{\dagger},\\
 && A_5(x^\mu,\pi R-y)   = -P'A_5 (x^\mu, \pi R+y)P'^{\dagger}. 
\end{eqnarray}
under the parity transformation of $Z_2'$. 
It should be noticed that 
 the signs of parities of $A_5$ are opposite to
 those of $A_\mu$.  
On the other hand, 
 two bulk Higgs fields, $H_i$ $(i=u,d)$, 
 which are ${\bf 10}$ representation
 of $SO(10)$, 
 transform as 
\begin{eqnarray}
\label{eta-s}
&&H_i(x, -y) = P H_i(x, y) ~~, ~~ 
H_i(x, \pi R-y) = P' H_i(x, \pi R + y) , 
\end{eqnarray}
under the parities.

The boundary conditions 
 are taken as 
 $P = \sigma_2\otimes I_5$ and 
 $P' = I_2\otimes \diag(1,1,1,-1,-1)$
 in the base
 of $SO(10)$, 
 which commute with the generators
 of the Georgi-Glashow $SU(5)\times U(1)_X$\cite{GG} and 
 the Pati-Salam $SU(4)_{PS}\times SU(2)_L\times SU(2)_R$ groups, 
 respectively\cite{ABC,HNOS}.
Seeing the zero modes of 
 gauge and Higgs fields in the bulk, 
 we can show that 
 the $SO(10)$ 
 gauge group is broken down to 
 $SU(3)_C\times SU(2)_L \times U(1)_Y \times U(1)_X$, 
 and the TD splitting is 
 realized.  
Besides, there appear two zero modes, 
 ${\bf (3,2)}$ and ${\bf (\overline{3},2)}$, from 
 $A_5$ as the Wilson line degrees of freedom.  
They are unwanted massless scalar 
 fields in the low energy. 
This is the reason why 
 many people have considered 
 $SO(10)$ GUT in not 5D but 6D, 
 where 
 the above 
 unwanted massless scalar fields 
 obtain KK masses and do not appear
 in the low energy\footnote{In Ref.\cite{ABC,HNOS}, 
 they take $T^2/(Z_2 \times Z_2')$ orbifold with
 the boundary condition as 
 $P_5=I_{10}$, $T_5 = \sigma_2\otimes I_5$ for 
 the 5th coordinate and 
 $P_6=I_{10}$, $T_6 = I_2\otimes \diag(1,1,1,-1,-1)$ for 
 the 6th coordinate. 
This boundary condition 
 can avoid 
 massless zero modes of   
 ${\bf (3,2)}$ and ${\bf (\overline{3},2)}$, from 
 $A_5$ and $A_6$.}. 
However, radiative correction can induce
 the mass of the Wilson line phases in general.
So we need to calculate 
 the one loop
 effective
 potential of the Wilson line phases  
 to study the vacuum structure 
 and estimate the radiatively induced mass 
 of the Wilson line phases   
 in 5D theory. 
The 5D theory has a merit of a simple
 anomaly cancellation comparing to
 the 6D theory\footnote{The anomaly cancellation is not 
 trivial in such a 6D theory,
 for examples, there should be accompanied 
 two {\bf 10} hypermultiplets with a $SO(10)$
 gauge multiplet, and a {\bf 10} hypermultiplets 
 with a {\bf 16} (or $\overline{\bf 16}$)
 in ${\mathcal N}=1$ SUSY ((1,0-SUSY)). }.

\section{The effective potential of $SO(10)$}

At first we show 
 the effective potential 
 induced from the gauge and 
 ghost contributions.
The adjoint representation, {\bf 45},  
 is divided by
 $P = \sigma_2\otimes I_5$ and 
 $P' = I_2\otimes \diag(1,1,1,-1,-1)$. 
$P$ and $P'$ break $SO(10)$ to 
 $SU(5)\times U(1)_X$ and 
 $SU(4)_{PS}\times SU(2)_L\times SU(2)_R$,
 respectively.
The adjoint representation
 is decomposed as 
\begin{eqnarray}
&&{\bf 45 = 24_0 + 1_0 + 10_4 + \overline{10}_{-4}}, \\
&&{\bf 45 = (15,1,1)+(1,3,1)+(1,1,3)+(6,2,2)},
\end{eqnarray}
under $P$ and $P'$, respectively.

The Wilson line degrees of freedom 
 are ${\bf (3,2)}$ and 
 ${\bf (\overline{3},2)}$ components of 
 $A_5$ in the base of $SU(3)_c\times SU(2)_L$. 
In the base of 
 flipped $SU(5)_F\times U(1)$\cite{flipped}, to which $SO(10)$ 
 is broken by the product of parities $PP'$, 
 the VEVs of them 
 can be set as 
\begin{equation}
  \VEV{A_5}
    ={1\over 2gR} \Ls
    \begin{array}{ccc|cc}
      0 & 0 & 0 & a & 0\\
      0 & 0 & 0 & 0 & b\\
      0 & 0 & 0 & 0 & 0\\ \cline{1-5}
      a & 0 & 0 & 0 & 0\\
      0 & b & 0 & 0 & 0
    \end{array}\Rs_0 
\end{equation}
by the residual global symmetry.

Since the adjoint representation is decomposed as 
${\bf 45 = 24 + 1 + 10 + \overline{10} }$
in terms of the flipped $SU(5)_F$, 
the eigenvalues of 
 $D_y(A_5)^2$ for a adjoint field $B$ 
 depending on 
 VEVs are calculated as
\begin{eqnarray}
&& \times{(n\pm a)^2\over R^2}, \;\;\;
 \times{(n\pm b)^2\over R^2}, \;\;\;
 2\times{(n\pm a/2)^2\over R^2}, \;\;\;
 2\times{(n\pm b/2)^2\over R^2}, \nonumber \\
&& 2\times{(n\pm (a-b)/2)^2\over R^2}, 
 2\times{(n\pm (a+b)/2)^2\over R^2}, \;\;\;
 2\times{(n\pm (a-1)/2)^2\over R^2}, \nonumber \\
&& 2\times{(n\pm (b-1)/2)^2\over R^2}, 
 2\times{(n\pm (a-b-1)/2)^2\over R^2}, \;\;\;
 2\times{(n\pm (a+b-1)/2)^2\over R^2}, 
\end{eqnarray}
where the eigenfunctions of $B$ are expanded by
 $\cos{ny\over R}$ and $\sin{ny\over R}$ 
 ( $\cos{(n+1/2)y\over R}$ and $\sin{(n+1/2)y\over R}$)
 for $PP'=+$($PP'=-$) components. 
According to the calculational method proposed 
 in Ref.\cite{HY-proof}, 
 we find the effective potential 
 induced from the gauge sector is given by
\begin{eqnarray}
V_{\eff}^{gauge}
&=& -{3\over2} C \sum_{n=1}^{\infty}{1\over n^5}
    [\cos(2\pi na)+\cos(2\pi nb)+2\cos(\pi na)+2\cos(\pi nb) \nonumber \\
&& +2\cos(\pi n(a-b))+2\cos(\pi n(a+b))+2\cos(\pi n(a-1)) \nonumber \\
&& +2\cos(\pi n(b-1))+2\cos(\pi n(a+b-1))+2\cos(\pi n(a-b-1))] .
\label{gauge-contribution}
\end{eqnarray}
where $C \equiv 3/(64\pi^7R^5)$.

Next let us show the contributions from 
 Higgs fields in the bulk. 
The {\bf 10} representation 
 is divided as 
\begin{eqnarray}
&&{\bf 10 = 5_2 + \overline{5}_{-2}}, \\
&&{\bf 10 = (6,1,1)+(1,2,2)},
\end{eqnarray}
under $P$ and $P'$, and
\begin{equation}
 {\bf 10 = 5 + \overline{5} }
\end{equation}
under $PP'$.
The eigenvalues of 
 $D_y(A_5)^2$ for a {\bf 10} representation field $B$ 
 depending on VEVs are 
\begin{eqnarray}
&& {(n\pm a/2)^2\over R^2}, \;\;\;
{(n\pm b/2)^2\over R^2}, \;\;\;
{(n\pm (a-1)/2)^2\over R^2}, \;\;\;
{(n\pm (b-1)/2)^2\over R^2}, 
\end{eqnarray}
where the eigenfunctions of $B$ are expanded by
 $\cos{ny\over R}$, $\sin{ny\over R}$, 
 $\cos{(n+1/2)y\over R}$ and $\sin{(n+1/2)y\over R}$
 depending on the parities, $P$ and $P'$. 
Thus, the effective potential 
 induced from the Higgs sector is given 
 by
\begin{eqnarray}
V_{\eff}^{Higgs}
&=& -C \sum_{n=1}^{\infty}{1\over n^5}
    [\cos(\pi na)+\cos(\pi nb)+\cos(\pi n(a-1))+\cos(\pi n(b-1))], 
\label{Higgs-contribution}
\end{eqnarray}
for one complex {\bf 10} Higgs scalar. 
When there are $N_{v,s}$ {\bf10} Higgs scalars
 in the bulk, 
 the effective potential from 
 the Higgs contributions
 is given by 
 $N_{v,s}\times$ Eq.(\ref{Higgs-contribution}).

When there is 
 a fermion multiplet ($\psi$) or 
 a spinor Higgs, {\bf 16}, in the bulk, 
 they contribute to the effective 
 potential. 
Under the parities, 
 they 
 transform as
\begin{eqnarray}
\label{eta-D}
&&\psi(x, -y) = \eta P \gamma^5 \psi(x, y) ~~, ~~
\psi(x, \pi R-y) = \eta' P' \gamma^5 \psi(x, \pi R + y) ~~,
\end{eqnarray}
where $\eta, \eta'=\pm$, and the effective potential 
 induced from these bulk fields depends on 
 the sign of the product, $\eta \eta'$.

The eigenvalues 
 of 
 $D_y(A_5)^2$ depending on VEVs 
 for a {\bf 16} representation field $B$
 with $\eta\eta'=\pm$ 
 are 
\begin{eqnarray}
&& {(n\pm a/2)^2\over R^2}, \;\;\;
{(n\pm b/2)^2\over R^2}, \;\;\;
{(n\pm (a-1)/2)^2\over R^2}, \;\;\;
{(n\pm (b-1)/2)^2\over R^2}, \nonumber \\
&&{(n\pm (a+b^{-1}_{+0})/2)^2\over R^2}, \;\;\;
{(n\pm (a-b^{-1}_{+0})/2)^2\over R^2},
\end{eqnarray}
where the eigenfunctions of $B$ are expanded by
 $\cos{ny\over R}$, $\sin{ny\over R}$, 
 $\cos{(n+1/2)y\over R}$ and $\sin{(n+1/2)y\over R}$
 depending on the parities, $P$ and $P'$. 
Thus the effective potential 
 induced from the {\bf 16} representation fields is given 
 by
\begin{eqnarray}
V_{\eff}^{{\bf16}^{(\pm)}}
&=& \frac{d^{(\pm)}}{2}C \sum_{n=1}^{\infty}{1\over n^5}
    [\cos(\pi na)+\cos(\pi nb)+\cos(\pi n(a-1))+\cos(\pi n(b-1)) 
\nonumber \\
&&    +\cos(\pi n(a+b^{-1}_{+0}))+\cos(\pi n(a-b^{-1}_{+0}))],
\label{16-contribution}
\end{eqnarray}
where $d$ denotes the number of degree of freedom, 
 for examples, 
 $+4$ for one Dirac fermion and $-2$ for one complex scalar. 
When there are $N_{s,s}^{(\pm)}$  
 {\bf 16} complex scalars and $N_{s,f}^{(\pm)}$ {\bf 16} 
 fermions with $\eta\eta'=\pm$ in the bulk, 
 the contribution to the effective potential from the {\bf16} fields 
 is given as 
 $\Ls4N_{s,f}^{(\pm)}-2N_{s,s}^{(\pm)}\Rs \times 
  V_{\eff}^{{\bf 16}^{(\pm)}}$.

The radiatively induced mass of the Wilson
 line phases at the symmetric point, $a=b=0$, 
 is easily calculated 
 by using the Riemann's zeta function, 
 $\zeta_R(d)=\sum_{n=1}^{\infty}{1\over n^d}$. 
The mass is given as
\beqn
m_{A_5}^2 
 &=& (gR)^2 \left.{\partial^2 V_{\eff} 
                   \over \partial a^2}\right|_{a=b=0}\nn\\
 &=& 2\pi g_4^2R^3 C \pi^2
   \Ll 
     \frac32 \Ls 10-6\times\frac34 \Rs 
    +\Ls 1-\frac34 \Rs N_{v,s} 
    +\Ls 1-3\times\frac34 \Rs\Ls -2N_{s,f}^{(+)}+N_{s,s}^{(+)} \Rs
\RR\nn\\&&\LL\hsp{2}
    +\Ls 3-\frac34 \Rs\Ls  -2N_{s,f}^{(-)}+N_{s,s}^{(-)} \Rs
   \Rl \zeta_R(3)\nn\\
 &=& {3\zeta_R(3)g_4^2 \over 128 \pi^4}
   \Ll 33+N_{v,s} 
    +5\Ls 2N_{s,f}^{(+)}-N_{s,s}^{(+)} \Rs
    -9\Ls  2N_{s,f}^{(-)}-N_{s,s}^{(-)} \Rs
   \Rl\times\Ls\frac{1}{R}\Rs^2.
\eeqn
where the 4D gauge coupling constant 
 $g_4$, which is defined as $g_4 = g/\sqrt{2\pi R}$, 
 is assumed to be ${\mathcal O}(1)$. 
So the zero modes, ${\bf (3,2)}$ and 
 ${\bf (\overline{3},2)}$, components of 
 $A_5$ in the base of $SU(3)_c\times SU(2)_L$
 obtain masses of ${\mathcal O}(1/R)$ from 
 the radiative corrections, 
 even for the minimal bulk content, that is, only 
 the gauge multiplet propagates in the bulk. 
When the compactification scale is around the GUT
 energy scale, the Wilson line phases obtain
 the heavy masses radiatively, and do not survive
 in the low energy.

This symmetric point, $a=b=0$, is 
 the global minimum 
 degenerated with the point $a=b=1$
 in the wide region of the parameter 
 space, $N_{v,s}$, $N_{s,f}^{(\pm)}$, $N_{s,s}^{(\pm)}$. 
Unless the value of $2N_{s,f}^{(-)}-N_{s,s}^{(-)}$ is
 larger than that of $2N_{s,f}^{(+)}-N_{s,s}^{(+)}$
 and/or $m_{A_5}^2<0$,
 where $(a,b)=(1,0)$ and $(0,1)$ points become 
 the global minimum, 
 the symmetric point is remaining as the global
 minimum. 
Even when the symmetric point is 
 a local minimum, 
 the tunnel transition from the symmetric 
 point to $(a,b)=(1,0)$ or $(0,1)$ points 
 is negligible as 
 long as $m_{A_5}^2\gg0$\cite{HHHK}.  
We can easily see the tunnel transition between 
 $a=b=0$ and $a=b=1$ is also negligibly small. 
So once the vacuum exists at $a=0$ in the early universe, 
 we can consider this color conserving 
 vacuum is stable.

The SUSY version of the 
 effective potentials are 
 easily obtained from those of the non-SUSY version 
 in the case of the 
 Scherk-Schwarz (SS) SUSY 
 breaking\cite{SS,SS2,SS3,SS4}. 
They are obtained by 
 adding a factor $(1-\cos(2\pi n\beta))$ in the summations 
 and modifying the coefficients depending on the 
 number of the degree of freedoms as 
 in Eq.(\ref{16-contribution})\cite{HY-proof}. 
The $\beta$ parameterizes the SS SUSY 
 breaking, which 
 should be taken as to induce 
 the soft mass of order 
 $\beta /R$\cite{HHHK}. 
Then the SUSY version of the effective potentials are 
 given by 
\begin{eqnarray}
V_{\eff}^{gauge}
&=& -{2} C \sum_{n=1}^{\infty}{1\over n^5}
    (1-\cos (2 \pi n \beta))
    [\cos(2\pi na)+\cos(2\pi nb)+2\cos(\pi na) \nonumber \\
&& +2\cos(\pi nb)+2\cos(\pi n(a-b))+2\cos(\pi n(a+b))
   +2\cos(\pi n(a-1)) \nonumber \\
&& +2\cos(\pi n(b-1))+2\cos(\pi n(a+b-1))+2\cos(\pi n(a-b-1))],
\label{gauge-contribution-susy} \\
V_{\eff}^{Higgs}
&=& {2C} \sum_{n=1}^{\infty}{1\over n^5}(1-\cos (2 \pi n \beta))
    \times \nonumber \\
&&    [\cos(\pi na)+\cos(\pi nb)+\cos(\pi n(a-1))+\cos(\pi n(b-1))], 
\label{Higgs-contribution-susy}\\
V_{\eff}^{\bf{16}^{(\pm)}}
&=& 2C \sum_{n=1}^{\infty}{1\over n^5}(1-\cos (2 \pi n \beta))
    [\cos(\pi na)+\cos(\pi nb)+\cos(\pi n(a-1))  \nonumber \\
&&   +\cos(\pi n(b-1))+\cos(\pi n(a+b^{-1}_{+0}))
     +\cos(\pi n(a-b^{-1}_{+0}))]
\label{16-contribution-susy}
\end{eqnarray}
for the gauge multiplet, one {\bf 10} hypermultiplet and 
 one {\bf 16} hypermultiplet with $\eta\eta'=\pm$, 
 respectively.

As in the non-SUSY case, 
 two points, $a=b=0$ and $a=b=1$, are degenerated and 
 become the global minimum
 when the number of {\bf 16} hypermultiplet with $\eta\eta'=-$ 
 is not so large, 
 and $m_{A_5}^2>0$. 
The tunnel transition between them
 is also negligible\cite{HHHK}.
The masses of the Wilson line phases at the symmetric point 
 are given, by using the approximation formula,
\begin{eqnarray}
\label{ap1}
&&\sum^{\infty}_{n=1} 
{\cos(n\xi) \over n^3}\simeq 
 \zeta(3)+{\xi^2 \over 2}\ln\xi-{3\over4}\xi^2, \\
&&\sum^{\infty}_{n=1} 
{\cos(n\xi) \over n^3}(-1)^n\simeq 
 -{3\over4}\zeta(3)+{\xi^2 \over 2}\ln2, 
\label{ap2}
\end{eqnarray}
for a small (positive) $\xi$, as 
\beqn
m_{A_5}^2 
 &=& (gR)^2 \left.{\partial^2 V_{\eff} 
                   \over \partial a^2}\right|_{a=b=0}\nn\\
 &=& 2\pi g_4^2R^3 C \pi^2
   \Ll 
     2 \Ls 10\times(-2\pi^2\beta^2\ln(2\pi\beta)+3\pi^2\beta^2)
    -6\times2\pi^2\beta^2\ln2 \Rs 
\RR\nn\\&&\LL\hsp{2}
    -2\Ls (-2\pi^2\beta^2\ln(2\pi\beta)+3\pi^2\beta^2)
          -2\pi^2\beta^2\ln2 \Rs N_v 
\RR\nn\\&&\LL\hsp{2}
    -2\Ls (-2\pi^2\beta^2\ln(2\pi\beta)+3\pi^2\beta^2)
           -3\times2\pi^2\beta^2\ln2 \Rs N_s^{(+)}
\RR\nn\\&&\LL\hsp{2}
    -2\Ls 3\times(-2\pi^2\beta^2\ln(2\pi\beta)+3\pi^2\beta^2) 
         -2\pi^2\beta^2\ln2 \Rs N_s^{(-)}
   \Rl \nn\\
 &=& {3g_4^2 \over 16 \pi^2}
   \Ll 
    (-20+2N_v+2N_s^{(+)}+6N_s^{(-)})\ln(2\pi\beta)
\RR\nn\\&&\LL\hsp{0.8}
   +(30-3N_v-3N_s^{(+)}-9N_s^{(-)})
\RR\nn\\&&\LL\hsp{0.8}
   +(-12+2N_v+6N_s^{(+)}+2N_s^{(-)})\ln2
   \Rl
   \times\Ls\frac{\beta}{R}\Rs^2, 
\eeqn
where $N_v$ and $N_s^{(\pm)}$ denote 
 the number of the {\bf10} hypermultiplets 
 and that of the {\bf10} hypermultiplets with $\eta\eta'=\pm$ 
 in the bulk, respectively.
Thus, 
 the scalar components of the zero modes, 
 ${\bf (3,2)}$ and ${\bf (\overline{3},2)}$, of 
 $A_5$ obtain masses of ${\mathcal O}(\beta/R)$ from 
 the radiative corrections.
As for the fermion components of 
 ${\bf (3,2)}$ and ${\bf (\overline{3},2)}$, 
 they also obtain masses of order $\beta/R$ via 
 the SS mechanism as the $\mu$-term generation%
\cite{SS3}
 in the gauge-Higgs unification scenario%
\cite{quasilocalize,Burdman:2002se}. 
It is because they are a part of a $SU(2)_R$ doublet. 
So that 
 all the component of the Wilson line phases 
 become massive, although 
 they might spoil the success 
 of the gauge coupling unification, 
 due to their small masses of ${\mathcal O}(\beta/R)$.

However, there is the possibility that 
 boundary localized kinetic terms can recover
 the gauge coupling unification. 
In general,
 boundary localized kinetic terms can exist 
 independently of bulk kinetic terms, 
 and affect the gauge coupling unification 
 in orbifold GUT scenarios. 
Although such contribution is often assumed to be 
 negligible, even order one boundary localized gauge couplings 
 may give the same order contribution from 
 one pair of ${\bf (3,2)}$ and ${\bf (\overline{3},2)}$ 
 chiral multiplet with a mass of order SUSY-breaking. 
For example, $g_2=1$ and $g_3=3/2$ on a boundary 
 induce the difference of the fine structure constants as 
\bequ
 \Delta\alpha^{-1}=4\pi\Ls\frac1{g_2^2}-\frac1{g_3^2}\Rs
                  =\frac{20}9\pi\sim7,
\eequ
while the pair contributes as
\bequ
 \Delta\alpha^{-1}\sim\frac1{2\pi}(b_2-b_3)\ln\Ls\frac{1/R}{\beta/R}\Rs
                  \sim5,
\eequ
where $b_a$ denotes the contribution to 
 the renormalization coefficient from the pair, $b_2=3, b_3=2$, 
 with $\beta\sim10^{-13}$.
So here we assume the gauge coupling unification is recovered 
 by the contribution from boundary localized kinetic terms.

\section{Summary and discussion}

In a 5D orbifold GUT, if $SO(10)$ is broken into 
 $SU(3)\times SU(2)\times U(1)^2$ 
 by the boundary conditions, 
 colored components of the 5th dimensional component 
 of the gauge field become massless at the tree level.
In order to give them mass, people have extended 
 the model to the 6D, or  
 introduced the additional bulk hypermultiplets 
 with the field developing non-vanishing VEVs 
 by hand. 
However, in this paper, 
 we have shown the radiative corrections 
 make the colored zero-modes acquire masses. 
In SUSY case, SS SUSY breaking makes the fermionic 
 partner of the zero-modes be massive, too. 
The boundary localized kinetic terms can 
 recover the gauge coupling 
 unification\footnote{
In a warped geometry, colored light modes may not 
 spoil the success of the gauge coupling unification 
 in the minimal SUSY $SU(5)$ model\cite{GCUwarped}. 
And a $SO(10)$ model is also considered in warped back ground%
 \cite{5dso10warped}.
}.

\vskip 1.5cm

\leftline{\bf Acknowledgments}

We would like to thank S. Komine for helpful discussions
in the early stage of this work. 
This work was supported in part by  Scientific Grants from 
 the Ministry of Education and Science, 
 Grant No.\ 14039207, 
 Grant No.\ 14046208, \ Grant No.\ 14740164 (N.H.), and 
 by a Grant-in-Aid for 
 the 21st Century COE ``Center for Diversity 
 and Universality in Physics'' (T.Y.).

\vskip 1cm

\def\jnl#1#2#3#4{{#1}{\bf #2} (#4) #3}

\def\Zphys{{\em Z.\ Phys.} }
\def\jssc{{\em J.\ Solid State Chem.\ }}
\def\jpsJ{{\em J.\ Phys.\ Soc.\ Japan }}
\def\ptps{{\em Prog.\ Theoret.\ Phys.\ Suppl.\ }}
\def\PTP{{\em Prog.\ Theoret.\ Phys.\  }}

\def\JMP{{\em J. Math.\ Phys.} }
\def\NPB{{\em Nucl.\ Phys.} B}
\def\NP{{\em Nucl.\ Phys.} }
\def\PLB{{\em Phys.\ Lett.} B}
\def\PL{{\em Phys.\ Lett.} }
\def\PRL{\em Phys.\ Rev.\ Lett. }
\def\PRB{{\em Phys.\ Rev.} B}
\def\PRD{{\em Phys.\ Rev.} D}
\def\PRe{{\em Phys.\ Rep.} }
\def\AP{{\em Ann.\ Phys.\ (N.Y.)} }
\def\RMP{{\
em Rev.\ Mod.\ Phys.} }
\def\ZPC{{\em Z.\ Phys.} C}
\def\SCI{\em Science}
\def\CMP{\em Comm.\ Math.\ Phys. }
\def\MPLA{{\em Mod.\ Phys.\ Lett.} A}
\def\IJMPA{{\em Int.\ J.\ Mod.\ Phys.} A}
\def\IJMPB{{\em Int.\ J.\ Mod.\ Phys.} B}
\def\EPJC{{\em Eur.\ Phys.\ J.} C}
\def\PR{{\em Phys.\ Rev.} }
\def\JHEP{{\em JHEP} }
\def\cmp{{\em Com.\ Math.\ Phys.}}
\def\JPA{{\em J.\  Phys.} A}
\def\CQG{\em Class.\ Quant.\ Grav. }
\def\ATMP{{\em Adv.\ Theoret.\ Math.\ Phys.} }
\def\ibid{{\em ibid.} }

\leftline{\bf References}

\renewenvironment{thebibliography}[1]
         {\begin{list}{[$\,$\arabic{enumi}$\,$]}  
         {\usecounter{enumi}\setlength{\parsep}{0pt}
          \setlength{\itemsep}{0pt}  \renewcommand{\baselinestretch}{1.2}
          \settowidth
         {\labelwidth}{#1 ~ ~}\sloppy}}{\end{list}}

\end{document}